\documentclass[aps,prb,twocolumn,floatfix,floats,showpacs,superscriptaddress,raggedbottom]{revtex4}
\usepackage{graphicx,latexsym}
\usepackage{dcolumn}
\usepackage{amsmath}
\pdfoutput=1
\begin{document}

\title{Correlated time-dependent transport through a 2D quantum structure}

\author{Vidar Gudmundsson}
\email{vidar@raunvis.hi.is}
\affiliation{Science Institute, University of Iceland, Dunhaga 3,
        IS-107 Reykjavik, Iceland}
\affiliation{Physics Division, National Center for Theoretical Sciences,
        PO Box 2-131, Hsinchu 30013, Taiwan}
\author{Chi-Shung Tang}
\email{cstang@nuu.edu.tw}
\affiliation{Department of Mechanical Engineering,
        National United University,
        1, Lienda, Miaoli 36003, Taiwan}
\author{Olafur Jonasson}
\affiliation{Science Institute, University of Iceland, Dunhaga 3,
        IS-107 Reykjavik, Iceland}
\author{Valeriu Moldoveanu}
\affiliation{National Institute of Materials Physics, P.O. Box MG-7,
        Bucharest-Magurele, Romania}
\author{Andrei Manolescu}
\affiliation{Reykjavik University, School of Science and Engineering,
        Kringlan 1, IS-103 Reykjavik, Iceland}

%

\begin{abstract}
We use a generalized master equation (GME) to describe the nonequilibrium
magnetotransport of interacting electrons through a broad finite quantum
wire with an embedded ring structure. The finite quantum wire is weakly
coupled to two broad leads acting as reservoirs of electrons.
The mutual Coulomb interaction of the electrons is described
using a configuration interaction method for the many-electron states of the
central system. We report some non-trivial interaction effects
both at the level of time-dependent filling of states and on the time-dependent transport.
We find that the Coulomb interaction in this non-trivial
geometry can enhance the correlation of electronic states in the system and
facilitate it's charging in certain circumstances in the weak coupling
limit appropriate for the GME.
In addition, we find oscillations in the current in the leads
due to the correlations oscillations caused by the switched-on
lead- system coupling. The oscillations are influenced and can be
enhanced by the external magnetic field and the Coulomb interaction.

\end{abstract}


\maketitle

%
%

\section{Introduction}
The Coulomb blocking of electrons entering nanostructures has been known for
quite some time and is explained by the magnitude of the direct part
of the repulsive Coulomb interaction energy in relation to the energy spectrum of the
nanosystem.\cite{Kouwenhouven97:01}
The experimental and theoretical studies of Coulomb effects on the mesoscopic transport
were commonly focused on steady-state regime of single or double quantum dots.
However, the increasing interest in fast dynamics at nanoscale and time-resolved detection
of electrons via a nearby detector strongly motivates theoretical investigations
of interacting time-dependent transport in complex systems.
The dynamical aspects of the Coulomb blocking have been investigated
by Kurth et al.\ in a one-dimensional lattice model using combination of non-equilibrium Green’s functions
and time dependent density-functional theory for the Coulomb interaction.\cite{Kurth09:11.3870}

In a recent work \cite{Moldoveanu10:IGME} we have also analyzed the transient currents
through an interacting two-dimensional quantum dot by solving the generalized master equation
for the matrix elements of the reduced density operator acting in the Fock space of interacting
many-electrons states of the dot. The GME scheme that we implement numerically takes into account
the geometrical details of the sample and leads (see Ref.\ \onlinecite{Moldoveanu09:073019}).
Our aim here is to apply the same method to more complex systems
in order to single out non-trivial Coulomb effects in the transient regime.
The system we consider is a parabolic quantum wire with an additional ring-shaped
confining potential. The embedded ring geometry imposes different localization properties
of the states with respect to the regions where the leads are attached.
This fact has important consequences on the time-dependent filling of the many-electron states.

Even though the transport and magnetic properties of quantum rings have fascinated
researchers for a long time new questions and results regarding the Aharonov-Bohm interference oscillations
have been catching attention.\cite{Camino09:155313,Camino07:155305,Ihnatsenka08:235304}
Dynamical effects have been studied in open or closed quantum rings.
Propagation of electron pulses in rings of finite width has been investigated by
Chaves et al.\cite{Chaves09:125331} and by Thorgillsson et al. \cite{Thorgilsson07:0708.0103} within scattering theory,
 and non-adiabatic current generation in a closed finite quantum ring in an external
magnetic field has been studied by integrating the Liouville-von Neumann equation
for the density operator in time.\cite{Gudmundsson03:161301} The system was perturbed
by a strong dipole or higher order multipolar electric electric field pulse and the
mutual Coulomb interaction between the electrons was included in a mean-field
DFT manner.

We use a non-Markovian version of the generalized master
equation\cite{Moldoveanu09:073019,Gudmundsson09:113007}
and treat the Coulomb interaction of the electrons exactly within a truncated many-electron
basis.\cite{Moldoveanu10:IGME} The GME formalism was originally proposed by Nakajima
and Zwanzig,\cite{Nakajima58:948,Zwanzig60:1338} and has more recently been applied to
study transport phenomena by several authors,
\cite{Timm08:195416,Harbola06:235309,Welack08:195315,Vaz08:012012} just to cite few.

The paper is organized as follows: Section II sets the notations and presents briefly the formalism,
Section III describes the quantum wire structure with the embedded ring, Section IV contains the results
and their discussion while Section V is left for conclusions.

\section{The GME and the Coulomb interaction}
In this Section we recall the main outlines of the GME method leading to the numerical results.
The time-dependent transport of noninteracting electrons with the generalized
master equation has been described in two publication for a lattice
model\cite{Moldoveanu09:073019} and a continuous model.\cite{Gudmundsson09:113007}
The non-interacting many-electron Hamiltonian of the coupled system
(i.e.\ the central sample, the leads and time-dependent coupling) reads as
\begin{equation}
      H(t)=\sum_a E_a d^\dagger_a d_a
      +\sum_{q,l=\mathrm{L,R}} \epsilon^l(q) c^\dagger_{ql} c_{ql}
      +H_\mathrm{T}(t),
\end{equation}
where the tunneling Hamiltonian $H_\mathrm{T}(t)=
H^\mathrm{L}_\mathrm{T}(t)+H^\mathrm{R}_\mathrm{T}(t)$
describes the coupling of the system to the left and right leads
\begin{equation}\label{H_tun}
      H^l_\mathrm{T}(t)=\chi^l(t)\sum_{q,a}
      \left\{T^l_{qa}c_{ql}^\dagger d_a + (T^l_{qa})^*d_a^\dagger c_{ql} \right\}.
\end{equation}
The creation and annihilation operators are associated to single particle states of the
disconnected subsystems. The coupling coefficient $T^l_{qa}$ of a single-electron state
$|q\rangle$ in the lead $l$ to a state $|a\rangle$ in the system is modelled as a non-local
overlap integral of the corresponding wave functions in the contact
regions of the system, $\Omega_S^l$, and the lead $l$,
$\Omega_l$\cite{Gudmundsson09:113007}
\begin{equation}
      T^l_{aq} = \int_{\Omega_S^l\times \Omega_l} d{\bf r}d{\bf r}'
      \left(\Psi^l_q ({\bf r}') \right)^*\Psi^S_a({\bf r})
      g^l_{aq} ({\bf r},{\bf r'})+h.c.
\label{T_aq}
\end{equation}
The function
\begin{align}
      g^l_{aq} ({\bf r},{\bf r'}) =
                   g_0^l&\exp{\left[-\delta_1^l(x-x')^2-\delta_2^l(y-y')^2\right]}\nonumber\\
                   &\exp{\left(\frac{-|E_a-\epsilon^l(q)|}{\Delta_E^l}\right)}.
\label{gl}
\end{align}
with ${\bf r}\in\Omega_\mathrm{S}^l$ and ${\bf r}'\in\Omega_l$
defines the `nonlocal overlap' and their affinity in energy.
The semi-infinite leads have the same parabolic
confinement as the finite quantum wire in the $y$-direction, perpendicular to the
transport direction $x$. The confinement is characterized by the energy scale
$\hbar\Omega_0$, The energy spectrum of the leads $\epsilon^l(q)$ is continuous, but
with clear subband structure. The effects of the external magnetic field
$\mathbf{B}=B\hat{\mathbf{z}}$ is present in the energy spectrum of the leads $\epsilon^l(q)$,
the spectrum of the system $E_a$, and in the wave functions of the leads and the system.

In order to describe the time-dependent transport when the system contains few electrons we select
the lowest $N_{\mathrm{SES}}$ single electron states (SES) of the central
system to construct a Fock space with $N_{\mathrm{MES}}=2^{N_{\mathrm{SES}}}$
many-electron states (MES). In the occupation representation basis such a state can be written as
\begin{equation}
      |\mu\rangle = |i^\mu_1,i^\mu_2,\dots,i^\mu_n,\dots\rangle ,
\end{equation}
where $i^\mu_n$ is the occupation of the $n-th$ single particle state of the isolated system.
$N_{\mathrm{SES}}$ is selected large enough that the chemical potentials of the leads
$\mu_l$ in equilibrium before the coupling at $t=0$ are smaller than the energy
of the highest SES, and ideally a further increase of $N_{\mathrm{SES}}$ should not change
the transport results of the calculations.

The Liouville-von Neumann equation describing the time-evolution of the total
system, central system and leads
\begin{equation}
      i\hbar\dot W(t)=[H(t),W(t)],\quad W(t<t_0)=\rho_\mathrm{L}\rho_\mathrm{R}\rho_\mathrm{S},
\label{L-vN}
\end{equation}
where the equilibrium density operator of the disconnected lead $l$ with
chemical potential $\mu_l$ is
\begin{equation}
      \rho_l=\frac{e^{-\beta (H_l-\mu_l N_l)}}{{\rm Tr}_l \{e^{-\beta(H_l-\mu_l N_l)}\}}.
\label{rho_l}
\end{equation}
is now projected on the central system by partial tracing operations with respect to
the operators of the leads. Defining the reduced density operator (RDO) of the central
system
\begin{equation}
      \rho(t)={\rm Tr}_\mathrm{L} {\rm Tr}_\mathrm{R} W(t),\quad \rho(t_0)=\rho_\mathrm{S},
\label{rdo}
\end{equation}
we obtain an integro-differential equation for the RDO, the generalized
master equation (GME)
\begin{eqnarray}
\nonumber
      {\dot\rho}(t)&=&-\frac{i}{\hbar}[H_\mathrm{S},\rho(t)]\\
      &-&\frac{1}{\hbar^2}\sum_{l=\mathrm{L,R}}\int dq\:\chi^l(t)
      ([{\cal T}^l,\Omega_{ql}(t)]+h.c.),
      \label{GME}
\end{eqnarray}
where two operators have been introduced to compactify the notation
\begin{eqnarray}
\nonumber
      &&\Omega_{ql}(t)=U_\mathrm{S}^\dagger (t) \int_{t_0}^tds\:\chi^l(s)
      \Pi_{ql}(s)e^{i((s-t)/\hbar )\varepsilon^l(q)}U_\mathrm{S}(t),\\\nonumber
      &&\Pi_{ql}(s)=U_\mathrm{S}(s)\left ({\cal T}^{l{\dagger}}
      \rho(s)(1-f^l)-\rho(s){\cal T}^{l{\dagger}}f^l\right )U_\mathrm{S}^\dagger(t),
\end{eqnarray}
with $U_\mathrm{S}(t)=e^{i(t/\hbar )H_\mathrm{S}}$,
and a scattering or coupling operator ${\cal T}$ acting in the Fock space of the system
\begin{eqnarray}
      {\cal T}^l(q)&=&\sum_{\alpha,\beta}{\cal T}_{\alpha\beta}^l(q)
      |{\bf \alpha}\rangle\langle {\bf \beta}|\\
      {\cal T}_{\alpha\beta}^l(q)&=&\sum_aT^l_{aq}\langle {\bf \alpha}
      |d_a^{\dagger}|{\bf \beta}\rangle.
\label{Toperator}
\end{eqnarray}
Here the kernel of the integro-differential equation has been obtained by
taking into account only second order processes with respect to the coupling
coefficients. It should though be kept in mind that the structure of the equation
implies higher order processes to infinite order.

In the derivation of the GME here only the coupling Hamiltonian is allowed to depend on time.
The possibility of the Hamiltonian of the central system $H_{\mathrm{S}}$
depending on time (describing a laser pulse for example) has been considered by
Amin et al.\cite{Amin09:103}
A DFT description of the mutual Coulomb interaction of the electrons in
the central system would require $H_{\mathrm{S}}$ to be time-dependent, and
it would also require a further reduction of the GME
introducing the reduced single-particle density matrix loosing some
many-electron correlation effects caused by the coupling of the system
to the leads.\cite{Li07:075114}

Our approach to solve this dilemma has been reported earlier,\cite{Moldoveanu10:IGME}
but here we shall briefly outline it for the case of the continuous model.
We choose to change the Hamiltonian of the central system
\begin{equation}
      H_{\mathrm{S}}=\sum_a E_a d^\dagger_a d_a
      +\frac{1}{2}\sum_{abcd}\langle ab|V_{\mathrm{Coul}}|cd\rangle
      d^\dagger_a d^\dagger_b d_d d_c
\label{HS_Coul}
\end{equation}
to include the time-independent Coulomb interaction term appropriate for
our many-electron formalism. The Coulomb potential is
\begin{equation}
      V_{\mathrm{Coul}}(\mathbf{r}-\mathbf{r}')=
      \frac{e^2}{\kappa\sqrt{(x-x')^2+(y-y')^2+\eta^2}},
\end{equation}
and the matrix elements are
\begin{equation}
      \langle ab|V_{\mathrm{Coul}}|cd\rangle =
      \int d{\mathbf{r}}\Psi^*_a(\mathbf{r})I_{bc}(\mathbf{r})\Psi_d(\mathbf{r})
\end{equation}
with
\begin{equation}
      I_{bc}(\mathbf{r})=\int d{\mathbf{r}'}\Psi^*_b(\mathbf{r}')
      V_{\mathrm{Coul}}(\mathbf{r}-\mathbf{r}')\Psi_c(\mathbf{r}') ,
\end{equation}
and $\eta$ is a small convergence parameter to be specified later.
Along the lines of approaches developed under the names of
``configuration interaction'' or ``exact numerical diagonalization'' we
diagonalize the new interacting Hamiltonian (\ref{HS_Coul}) in the MES
basis of the non-interacting system $\{|\mu\rangle\}$ in the entire
Fock space built from the $N_{\mathrm{SES}}$ SES states, including all
sectors containing zero electrons (the vacuum state) to $N_{\mathrm{SES}}$
electrons, since we are dealing with an open system with variable number
of electrons. The diagonalization yields a new basis of interacting
MES $\{|\mu)\}$ connected to the non-interacting one by a unitary
transformation
\begin{equation}
      |\mu) = \sum_{\alpha} {\cal V}_{\mu\alpha}|\alpha\rangle  ,
\end{equation}
supplied by the diagonalization. Here we need to keep in mind that ${\cal V}$
will be represented by an $N_{\mathrm{MES}}\times N_{\mathrm{MES}}$ matrix in
numerical calculations. An inspection of the structure of the
non-interacting GME (\ref{GME}) reveals that the equation can be also
be transformed to the interacting basis $\{|\mu)\}$ by the unitary transformation.
Thus, in a numerical calculation a basis transformation of the many-electron
coupling matrix $\tilde{{\cal T}}^l(q) = {\cal V}^\dagger{{\cal T}}^l(q){\cal V}$
(\ref{Toperator}) and the insertion of the diagonalized matrix representation of
the interacting $H_\mathrm{S}$ in the GME (\ref{GME}) will give the RDO in the
interacting MES basis $\tilde{\rho} = {\cal V}^\dagger\rho{\cal V}$.
As all measurable quantities are in the end expressed as a partial trace with
respect to operators of the central system expectation values can be calculated in
the new basis and for the same reason in the non-interacting case we can again
obtain the mean value of the left or right current directly from the transformed
GME.

In our earlier publication\cite{Gudmundsson09:113007} neglecting the
Coulomb interaction it was clear that we could only effectively describe
the time-dependent transport through systems with up to 7 or 8 SES considered
relevant for the currents, situated in and around the window
of the chemical potentials of the two leads. This limitation was imposed by the
complex bandstructure of the energy spectrum in the broad leads employed in
the calculations. Here, we may have to include more SESs in order to describe
reasonably the interaction of the electrons in the central system. In order to
accomplish this we have to resort to a more refined truncation procedure than
we used for the non-interacting system: The unitary transformation can not be
truncated and has to include the $N_\mathrm{MES}$ states constructed initially
from the $N_\mathrm{SES}$ SESs. In the numerical calculations here we will employ
12 SES leading to 4096 MES. The unitary transformation of ${\cal T}^l(q)$ is
thus CPU-time intensive for all the $q$-values necessary for the leads, but it has
only to be performed once. After that it is possible to deploy a second truncation
to the GME by keeping only the $N'_\mathrm{MES}<<N_\mathrm{MES}$ MES with lowest
energy. Typically, for the parameters that we will select for the numerical calculations
here we need only $N'_\mathrm{MES}=32$, but this cutoff is very system dependent.

\section{Embedded quantum structure}
\begin{figure}[htbq]
      \includegraphics[width=0.42\textwidth,angle=0]{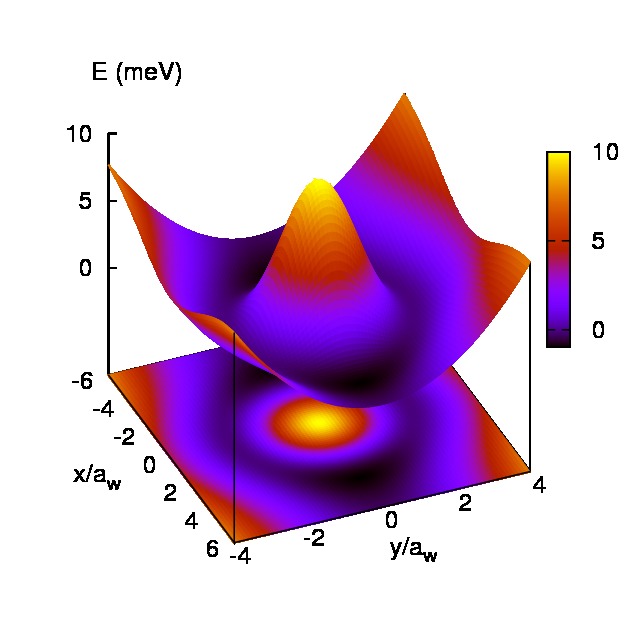}
      \caption{(Color online) The potential defining the quantum ring
               (\ref{V_QR}) embedded in the finite quantum wire at
               $B=1.0$ T $V_g=1.0$ meV, and $a_w=23.87$ nm.}
      \label{System}
\end{figure}

Here we will use the GME to analyze time-dependent transport of electrons
through a short but broad quantum wire of length $L_x=300$ nm, with an
embedded quantum ring. The parabolic confinement of the quantum wire is
characterized by the energy scale $\hbar\Omega_0 = 1.0$ meV, and we assume
GaAs parameters with $m^*=0.067m_e$ and $\kappa = 12.4$.
The embedded ring is represented by the potential
\begin{equation}
      V_\mathrm{QR}(\mathbf{r})=\sum_{i=1}^{2}V_i
      \exp{[-(\beta_{xi}x)^2-(\beta_{yi}y)^2]}+V_g,
\label{V_QR}
\end{equation}
with $V_1=-4.0$ meV, $V_2=+14.0$ meV, $\beta_{x1}=1.09\times 10^{-2}$ nm$^{-1}$,
$\beta_{y1}=3.46\times 10^{-4}$ nm$^{-1}$, $\beta_{x2}=1.09\times 10^{-2}$ nm$^{-1}$,
and $\beta_{y2}=2.83\times 10^{-2}$ nm$^{-1}$. The parameter $V_g$ can be thought of as
a gate voltage. We use it to position the chemical potential of the right lead
$\mu_R$ at a similar place in the energy spectrum of the SESs for the two values
of the magnetic field investigated here. (For $B=1.0$ T we use $V_g=1.0$ meV and for
$B=0.5$ T we have $V_g=1.2$ meV). Figure (\ref{System}) shows the ring
embedded in the quantum wire with the spatial coordinates scaled by the
magnetic length modified by the parabolic confinement $a_w=\sqrt{\hbar/(m^*\Omega_w)}$,
with $\Omega_w^2=\Omega_0^2+\omega_c^2$ at $B=1.0$ T, where the cyclotron frequency
is $\omega_c=eB/(m^*c)$. At $B=1.0$ T $a_w=23.87$ nm.
In the following calculations we use for the coupling (\ref{gl})
$\delta_1^l=4.39\times 10^{-4}$ nm$^{-2}$,
$\Delta_E^l=0.5$ meV, and $g_0^la_w^{3/2}=30$ meV or 40 meV.

\begin{figure}[htbq]
      \includegraphics[width=0.42\textwidth,angle=0]{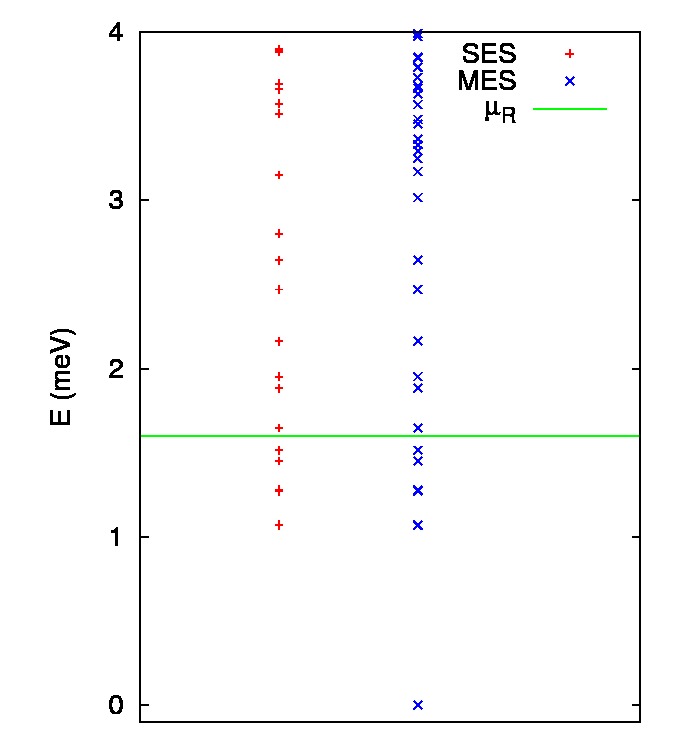}
      \caption{(Color online) The energy spectra for the single electron states
              (SESs) and the many-electron states (MESs) built from the lowest
              12 SESs. The solid green line indicates the chemical potential
              of the right lead $\mu_R=1.6$ meV. $B=1.0$ T, $V_g=1.0$ meV, $L_x=300$ nm.}
      \label{E_SES_MES}
\end{figure}

The energy spectra for the closed system of a quantum wire with an
embedded ring are shown for the SESs and the MESs in Figure \ref{E_SES_MES}
with the chemical potential of the right lead to be used in the following dynamical
calculations $\mu_R$ indicated.
As will be evident from the probabilities of the SESs the two lowest SES states
are almost degenerate, so below $\mu_R$ there are 6 SESs. In the dynamical
calculation to follow we will use the 12 lowest SESs to build the relevant
MESs. Besides the vacuum MES at zero energy we thus recognize the 12 SESs
again as MESs occupied by only one electron each. Above the energy of the
highest SES accounted for we have a relatively dense spectrum of MESs
occupied by 2 electrons. Due to the strong Coulomb interaction the lowest
MES occupied by 3 electrons is located well above the highest energy
shown in Fig.\ \ref{E_SES_MES}.

\begin{figure}[htbq]
      \includegraphics[width=0.15\textwidth,angle=0,viewport=25 45 210 210,clip]{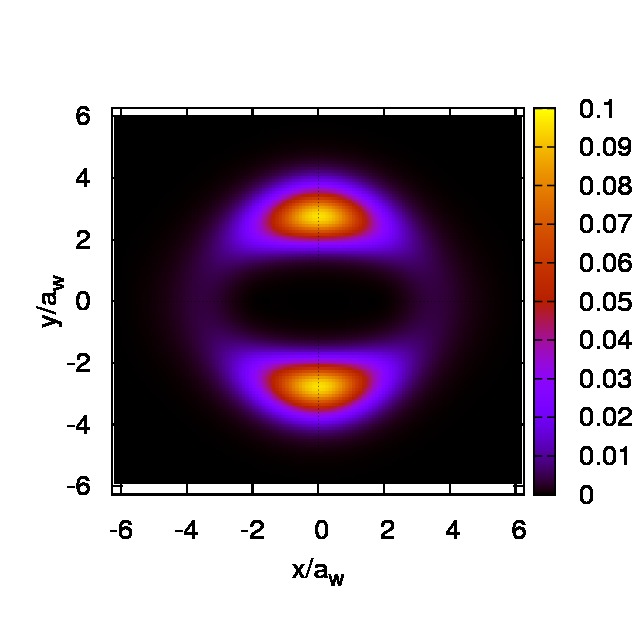}
      \includegraphics[width=0.15\textwidth,angle=0,viewport=25 45 210 210,clip]{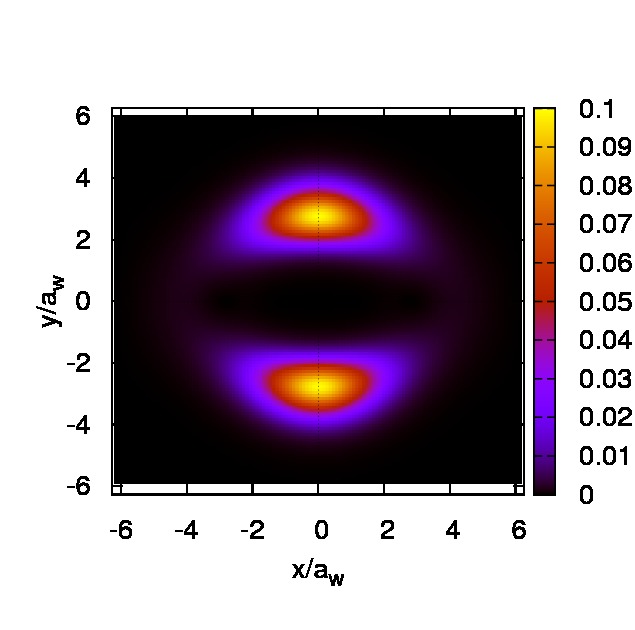}
      \includegraphics[width=0.15\textwidth,angle=0,viewport=25 45 210 210,clip]{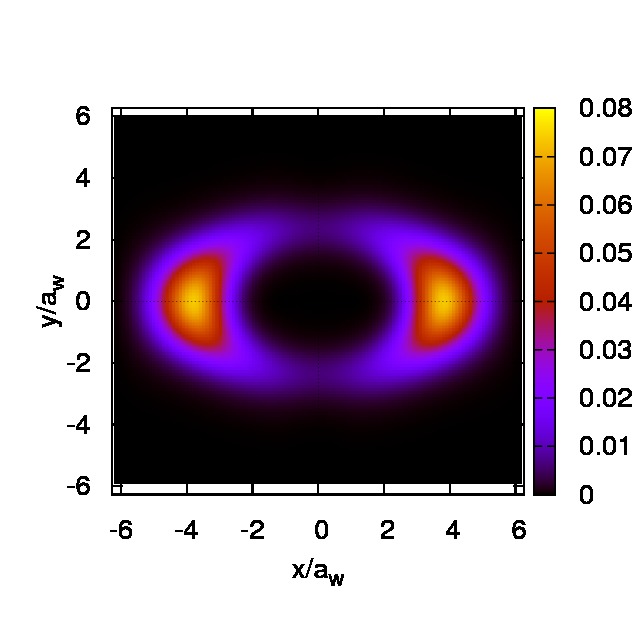}\\
      \includegraphics[width=0.15\textwidth,angle=0,viewport=25 45 210 210,clip]{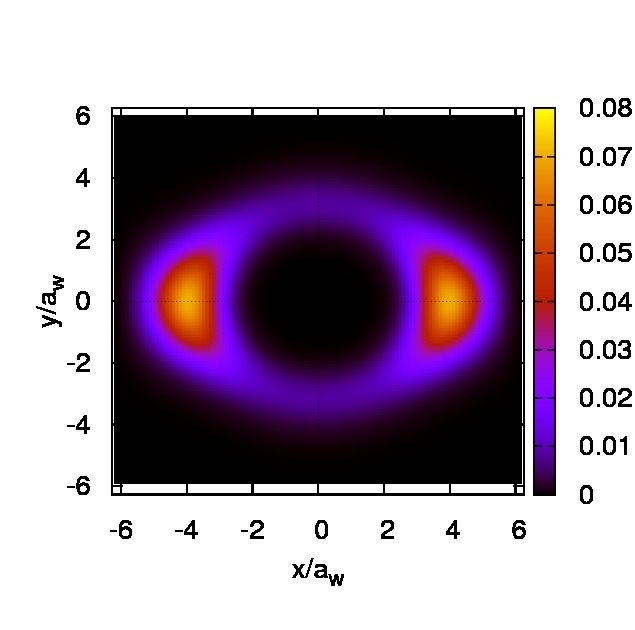}
      \includegraphics[width=0.15\textwidth,angle=0,viewport=25 45 210 210,clip]{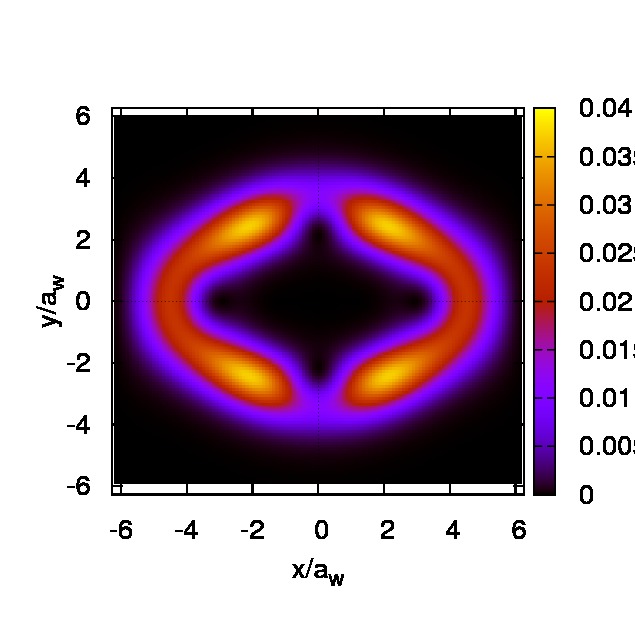}
      \includegraphics[width=0.15\textwidth,angle=0,viewport=25 45 210 210,clip]{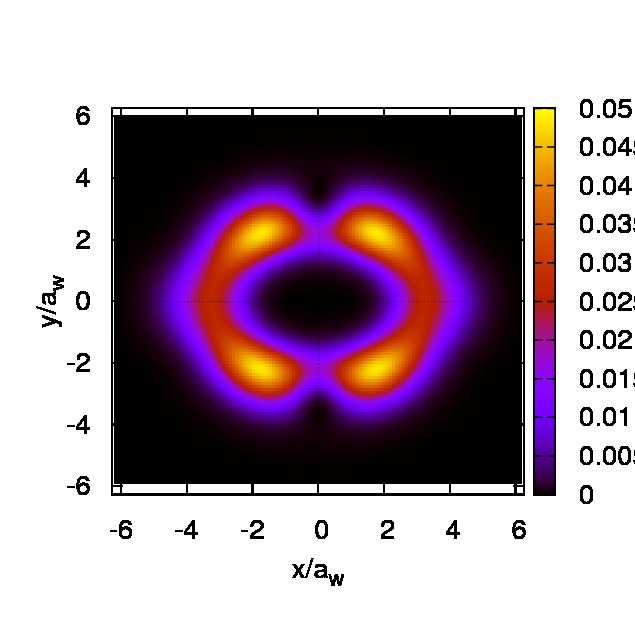}\\
      \includegraphics[width=0.15\textwidth,angle=0,viewport=25 45 210 210,clip]{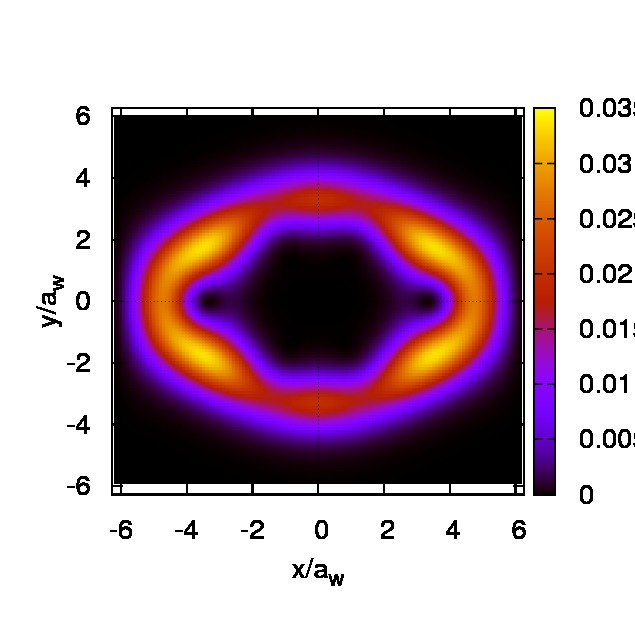}
      \includegraphics[width=0.15\textwidth,angle=0,viewport=25 45 210 210,clip]{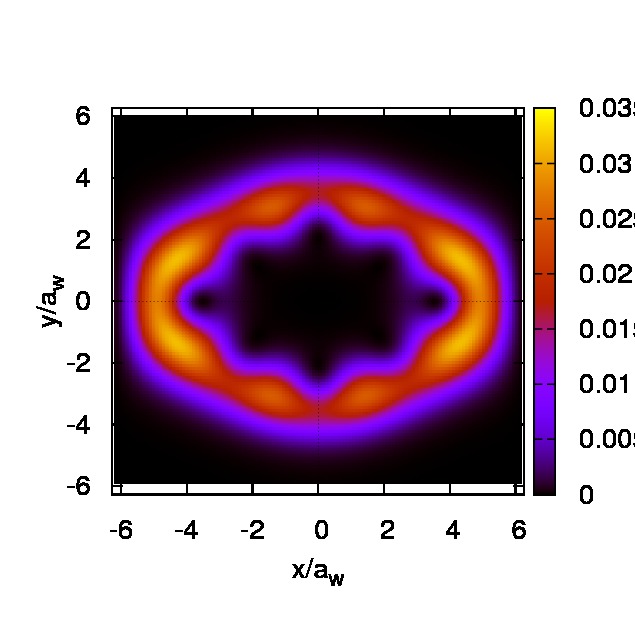}
      \includegraphics[width=0.15\textwidth,angle=0,viewport=25 45 210 210,clip]{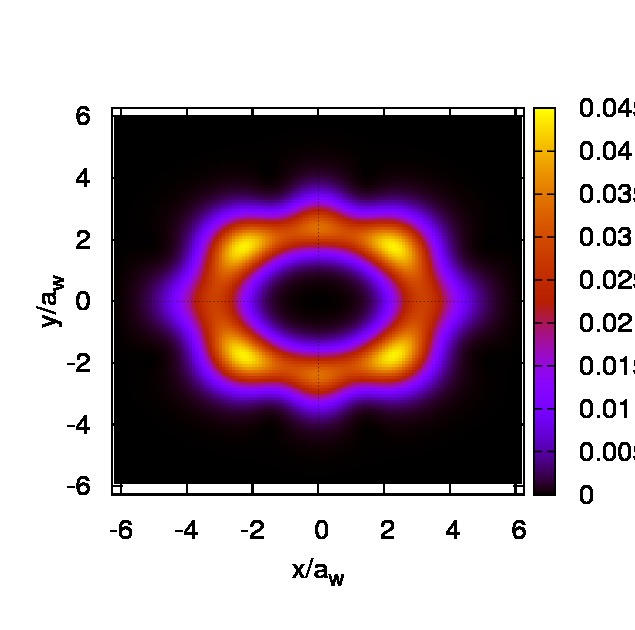}\\
      \includegraphics[width=0.15\textwidth,angle=0,viewport=25 12 210 210,clip]{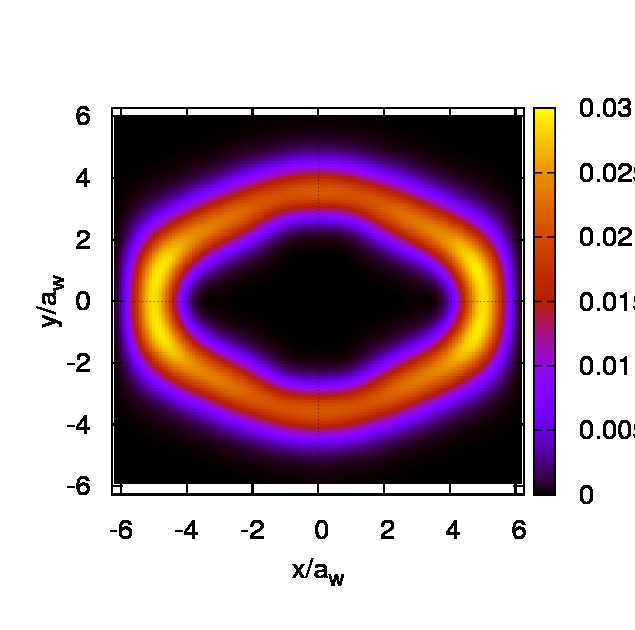}
      \includegraphics[width=0.15\textwidth,angle=0,viewport=25 12 210 210,clip]{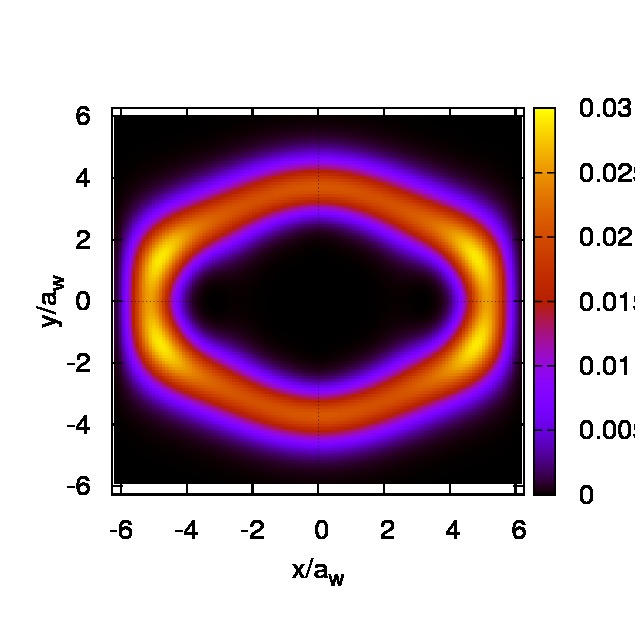}
      \includegraphics[width=0.15\textwidth,angle=0,viewport=25 12 210 210,clip]{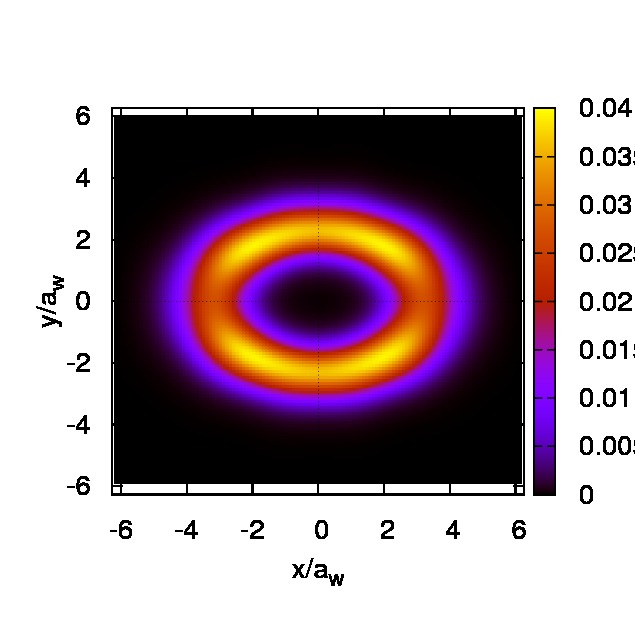}
      \caption{(Color online) The probability density of the single-electron eigenstates
               of the system labeled by $a$ in numerical order with $a=1$
               at the top left and $a=12$ at the bottom right. $B=1.0$ T, $L_x=300$ nm.}
      \label{Wf_Hr}
\end{figure}

The value of $\mu_R=1.6$ meV is selected such that below it there are both SESs
localized away from the contact region and states with a strong weight
in that region. The probability for the SESs is displayed in Figure \ref{Wf_Hr}.
%

\section{Dynamical transport properties}
In a recent paper \cite{Moldoveanu10:IGME} among other things we demonstrated a
dynamical Coulomb blocking effect in a small system where all the relevant SESs
are extended. There, the mutual Coulomb interaction between the electrons in the
system prevents the entrance of further electrons until the bias is high enough
and the occupation of the system in the steady state regime shows the well known
Coulomb steps as a function of the bias between the left and right
leads. The time-dependence of the contact functions $\chi^l$ is described by
\begin{equation}
      \chi^{\mathrm{L,R}}(t)=\left(1-\frac{2}{e^{\alpha^{\mathrm{L,R}} t}+1}\right) ,
\label{chi}
\end{equation}
with $\alpha^l = 1.0$ ps$^{-1}$. We fix the temperature of the reservoirs at
$T=0.5$ K.

\subsection{Enhanced occupation by the Coulomb interaction}
\begin{figure}[htbq]
      \includegraphics[width=0.42\textwidth,angle=0]{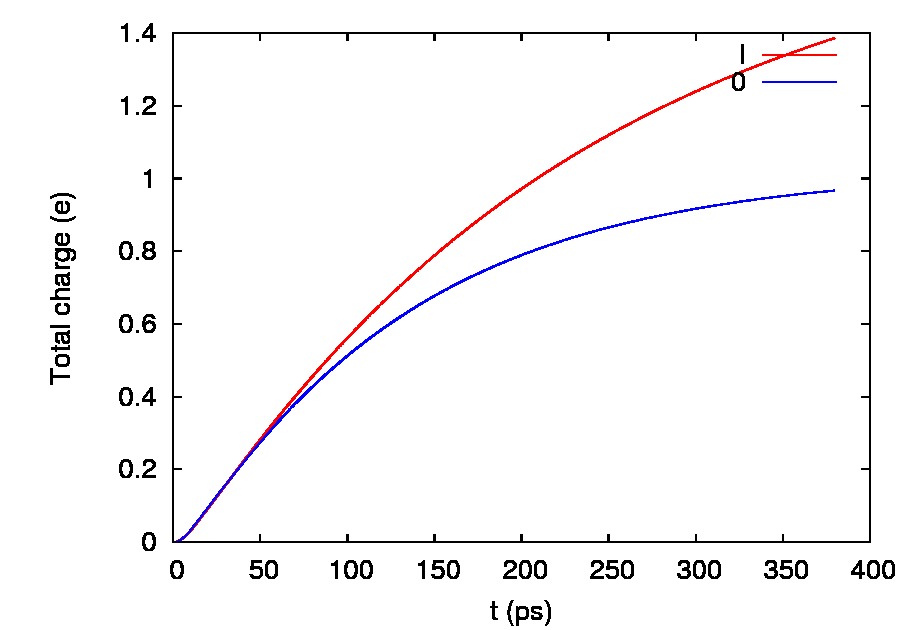}
      \caption{(Color online) The total charge of the non-interacting (0) and the
               interacting system (I) as function of time. $\Delta\mu=0.1$ meV,
               $B=1.0$ T, $V_g=1.0$ meV, $L_x=300$ nm, and $g_0^la_w^{3/2}=30$ meV.}
      \label{TC_CIO}
\end{figure}

\begin{figure}[htbq]
      \includegraphics[width=0.42\textwidth,angle=0]{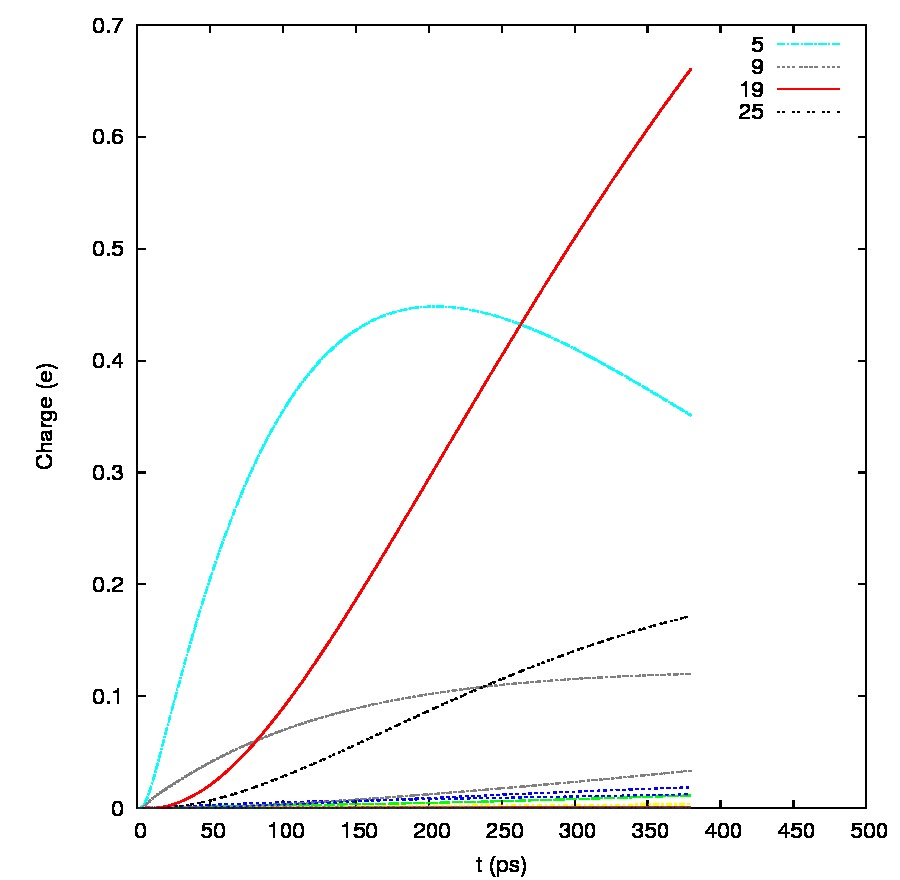}
      \caption{(Color online) The occupation of the interacting MES state $|\mu )$
               as a function of time. $\Delta\mu=0.1$ meV, $B=1.0$ T, $V_g=1.0$ meV,
               $L_x=300$ nm, and $g_0^la_w^{3/2}=30$ meV.}
      \label{TC_muI}
\end{figure}

\begin{figure}[htbq]
      \includegraphics[width=0.42\textwidth,angle=0]{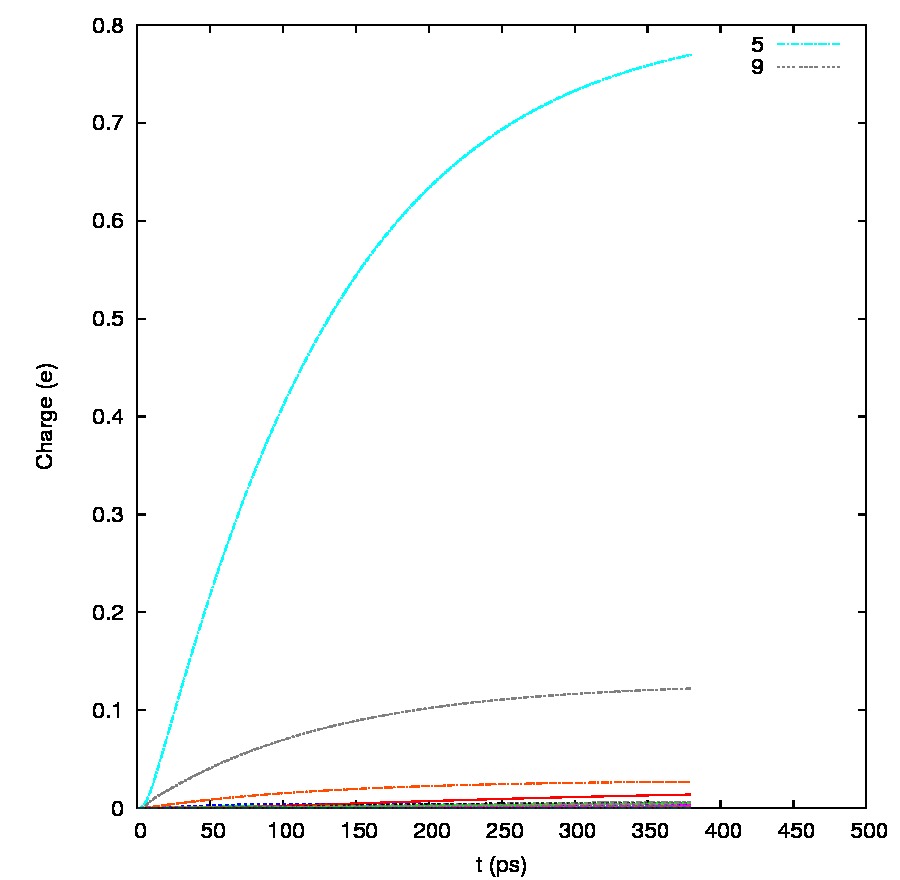}
      \caption{(Color online) The occupation of the non-interacting MES state $|\mu )$
               as a function of time. $\Delta\mu=0.1$ meV, $V_g=1.0$ meV,
               $B=1.0$ T, $L_x=300$ nm, and $g_0^la_w^{3/2}=30$ meV.}
      \label{TC_mu0}
\end{figure}

Here, in a system where not all the states of the system are extended we
will show that the Coulomb interaction can bring about a totally different
dynamical effect: Fig.\ \ref{TC_CIO} shows that even for a small bias
$\Delta\mu = \mu_L-\mu_R=0.1$ meV only one electron seems to be able to enter
the initially empty system in the absence of the Coulomb interaction, but
the Coulomb interaction seems to facilitate the entrance of
the second electron into the system.
A glance at Figures \ref{TC_muI} and \ref{TC_mu0} showing the occupation of the
MESs $|\mu )$ in case of the interacting and the non-interacting system, respectively,
indicates a very different charging effect for the two cases.
In the case of the non-interacting system the one-electron MES $|5)=|000100000000\rangle$
is occupied up to 78\% after $t=380$ ps and the state $|9)=|000000001000\rangle$
carries 12\% with the rest distributed to several states.
For the interacting system the one-electron state
$|5)=|000100000000\rangle$ is again initially occupied
with a slight occupation of $|9)=|129\rangle =|000000010000\rangle$, but soon they loose to
the two-electron MES $|19)$ and $|25)$ that take over.
The state $|19)$ is a two-electron state with
the main contributions from $|000110000000\rangle$,  $|100010000000\rangle$,
$|001100000000\rangle$, $|010100000000\rangle$, and $|101000000000\rangle$.
$|25)$ is also a two-electron MES with the main contributions
from $|100000100000\rangle$, $|100001000000\rangle$, $|010000010000\rangle$,
$|000100100000\rangle$, and $|000101000000\rangle$.

\begin{figure}[htbq]
      \includegraphics[width=0.42\textwidth,angle=0]{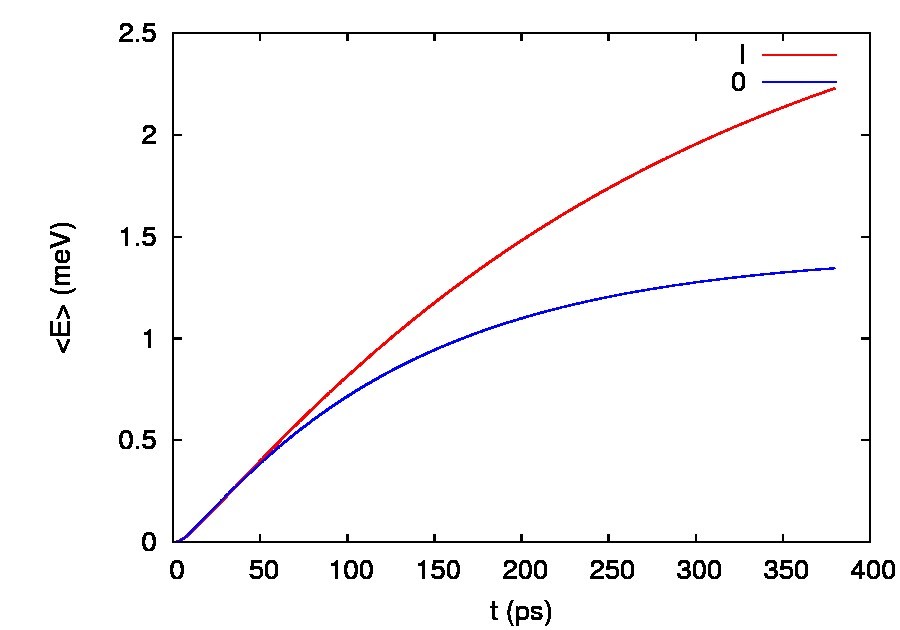}
      \caption{(Color online) The mean energy of the non-interacting and interacting
               system as function of time. $\Delta\mu=0.1$ meV, $B=1.0$ T, $V_g=1.0$ meV,
               $L_x=300$ nm, and $g_0^la_w^{3/2}=30$ meV.}
      \label{E_meanCIO}
\end{figure}

The energy of the state $|19)$ is $3.36$ meV and the mean energy for the same range
of time seen in Fig.\ \ref{E_meanCIO} is higher for the interacting case.
We assume the leads are in equilibrium before the coupling to the wire
at $t=0$ at a temperature of $0.5$ K corresponding to 0.043 meV.
The energy $3.36$ meV of $|19)$ is valid for the case of exactly
two electrons in the state, here we can only explain the occupation
of this state well above $\mu_L$ with the fact that it is only
partially occupied. In other words the SES elements which build the
MES $|19)$ are only partially occupied. This corresponds well with the
values of the mean energy in Fig.\ \ref{E_meanCIO}.

In the non-interacting case one electron can enter the system and it occupies the
state $|5)=|000100000000\rangle$ just below $\mu_R$. This state is the lowest
state with high weight in the contact region, below which only states
($a=1$, $a=2$) exist with more weight away from the region of contacts, in the
sides of the quantum ring, as Fig.\ \ref{Wf_Hr} confirms.
(The state $a=3$, or $|001000000000\rangle$, is nearly degenerate with
$|5)$ but does not participate in the transport to large extent
for the non-interacting case).
The Coulomb interaction couples together these two different types of states
and facilities thus the occupation of two-electron states with one of the
electrons in a low energy SES state with poor coupling to the contacts.
Similar phenomena is observed at $B=0.5$ T if the chemical potential
in the right lead is placed in a corresponding location with respect to the
12 non-interacting SESs used in the calculation by varying $V_g$.

\begin{figure}[htbq]
      \includegraphics[width=0.42\textwidth,angle=0]{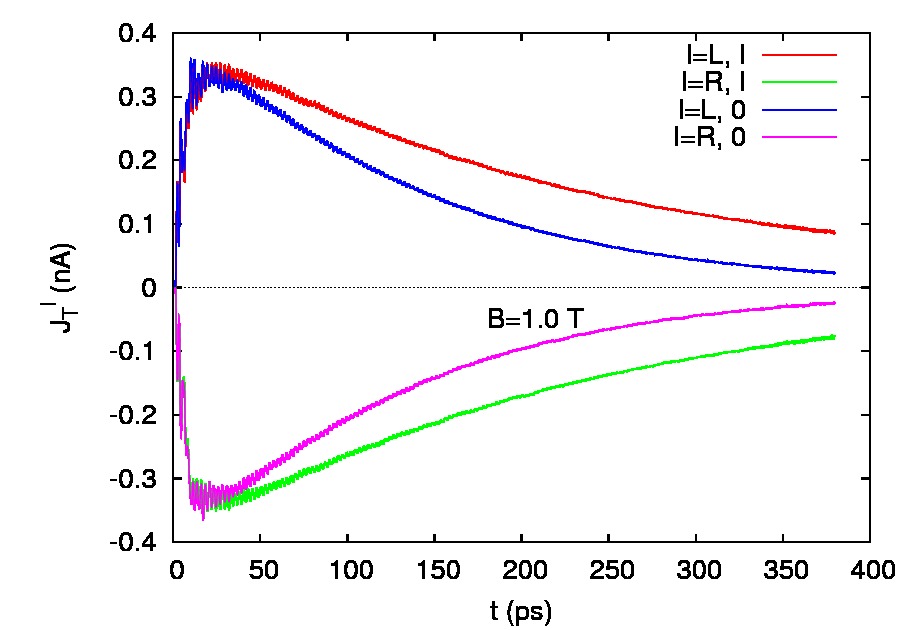}
      \caption{(Color online) The current in the left and right leads as a function of
               time for the non-interacting and the interacting system. $\Delta\mu=0.1$ meV,
               $B=1.0$ T, $V_g=1.0$ meV, $L_x=300$ nm, and $g_0^la_w^{3/2}=30$ meV.}
      \label{Current_CI0}
\end{figure}

The current in the left and right leads displayed in Fig.\ \ref{Current_CI0}
shows that neither the interacting nor the non-interacting systems have reached
a steady state in the 380 ps shown. Both leads are still supplying charge to
the system, but in the case of the interacting system the time constants are
clearly longer for the charging process that is enhanced by the Coulomb interaction.

\begin{figure}[htbq]
      \includegraphics[width=0.23\textwidth,angle=0,viewport=12 12 210 210,clip]{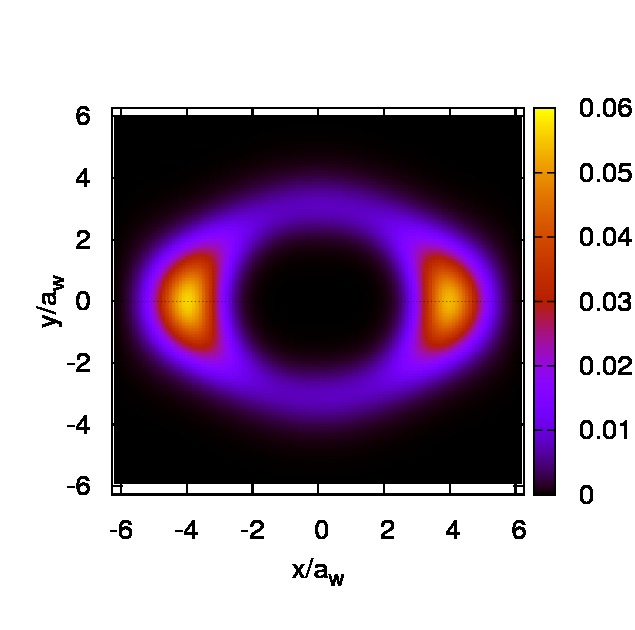}
      \includegraphics[width=0.23\textwidth,angle=0,viewport=12 12 210 210,clip]{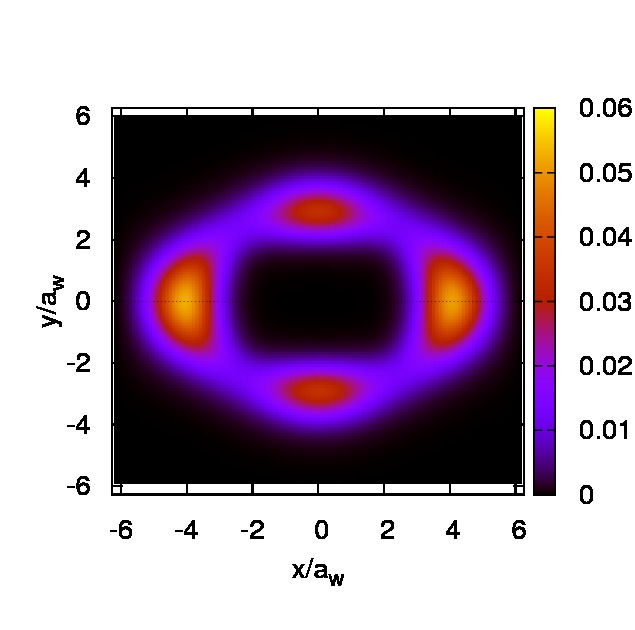}
      \caption{(Color online) The total many-electron charge density for the
               non-interacting (left panel) and interacting system (right panel) at
               $t=380$ ps. $\Delta\mu=0.1$ meV, $B=1.0$ T, $V_g=1.0$ meV, $L_x=300$ nm,
               and $g_0^la_w^{3/2}=30$ meV.}
      \label{Qnn_CIO}
\end{figure}

The many-electron charge distributions compared in Fig.\ \ref{Qnn_CIO}
for the non-interacting and the interacting system at $t=380$ ps
confirm this observation and reminds us that the two-electron state
in the right panel has a relatively low interaction energy due to the
reduced overlap of states with high probability in the contact region and
states with high probability at the other sides of the ring.

\begin{figure}[htbq]
      \includegraphics[width=0.23\textwidth,angle=0,viewport=12 12 250 210,clip]{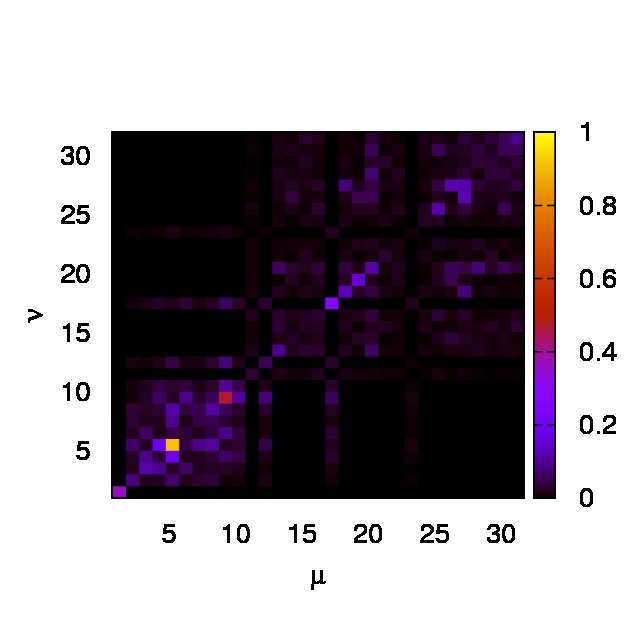}
      \includegraphics[width=0.23\textwidth,angle=0,viewport=12 12 250 210,clip]{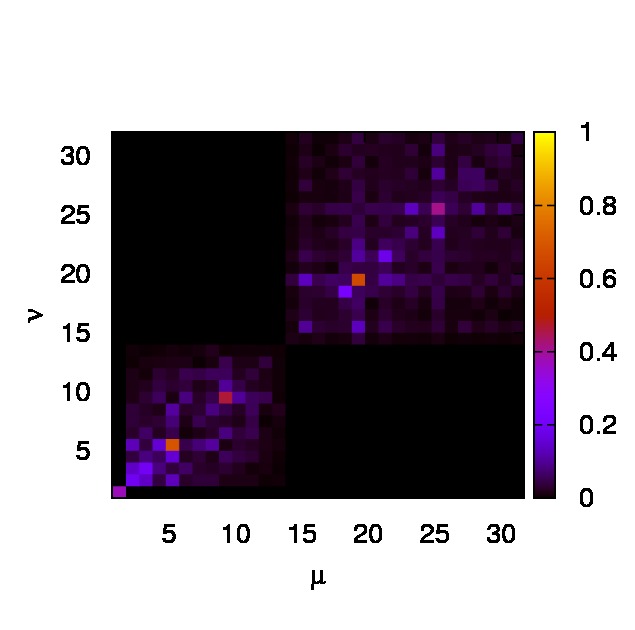}
      \caption{(Color online) The reduced density matrix $|\rho_{\mu\nu}|^{0.36}$ for the
               non-interacting (left panel) and interacting system (right panel) at
               $t=380$ ps. $\Delta\mu=0.1$ meV, $B=1.0$ T, $V_g=1.0$ meV, $L_x=300$ nm,
               and $g_0^la_w^{3/2}=30$ meV.
               The power $0.36$ is chosen to make the smaller off-diagonal elements visible on
               scale needed for the larger diagonal elements.}
      \label{rdo_122}
\end{figure}

In Fig.\ \ref{rdo_122} we compare the truncated RDO $\rho_{\mu\nu}$ for the case of
the non-interacting and the interacting system at $t=380$ ps. On the diagonal
in the left panel we see again as in Fig.\ \ref{TC_mu0} that for the non-interacting
case only state $|5)$ has considerable occupation, while the vacuum state $|1)$ is
loosing its initial high value and state $|9)$ is gaining some weight. So, only one-electron
states are occupied here. In the case for the interacting system the right panel
of Fig.\ \ref{rdo_122} shows as Fig.\ \ref{TC_muI} a strong emergent occupation of $|19)$,
a two electron state. Figure \ref{E_SES_MES} revealed a small energy gap between the
single-electron states and the many-electron states of the interacting system.
The GME-formalism excludes any correlation between MESs with a different number of
electrons. A manifestation of this can be seen in the right panel of Fig.\ \ref{rdo_122}
where only vanishing off-diagonal elements can be found in the upper left and the
lower right rectangles correlating one- and two-electron states.
(The one-electron states being $\mu = 1,2,\cdots ,12$). For the non-interacting
case in the left panel of Fig.\ \ref{rdo_122} this separation is not as clear cut since
there the regions of one- and two electrons states overlap slightly.
More importantly, Fig.\ \ref{rdo_122} reveals a non-vanishing correlation between all
two-electron states that gain any occupation in the system.
Initially for both cases only the vacuum state $|1)$ was occupied, so clearly the coupling
to the leads brings about correlation of the electrons in the system, and in addition, the
Coulomb interaction strongly influences this correlation.

\subsection{Current oscillations}
In this subsection we will compare the total current in the leads for two different values of the
magnetic field, i.e.\ at $0.5$ T and $1.0$ T, and observe how the current changes
as the bias is increased. We find that the current exhibits smooth oscillations
with a period of several picoseconds for the higher magnetic field as the
bias is increased. Here we should mention right away that we are not describing the
small oscillations seen in Fig.\ \ref{Current_CI0} at a shorter time scale, that are
caused by an interference of the coupling to different subbands of the leads.
(In the case of a one-dimensional lead like has been used in the lattice version
of the GME-model these oscillations do not appear).\cite{Moldoveanu09:073019}

\begin{figure}[htbq]
      \includegraphics[width=0.42\textwidth,angle=0]{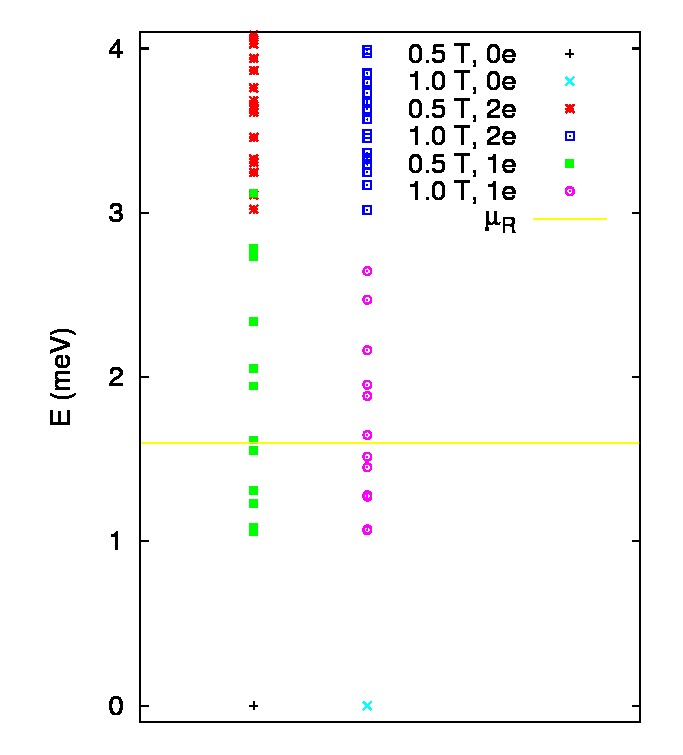}
      \caption{(Color online) The energy spectra for the lowest 32 many-electron states (MESs)
              built from the lowest 12 SESs compared for two values of the magnetic field.
              The solid horizontal yellow line indicates the chemical potential
              of the right lead. Different symbols are used for one-  and two electron states.
              $\mu_R=1.6$ meV. $L_x=300$ nm. At $B=1.0$ T $V_g=1.0$ meV, and at
              $B=0.5$ T $V_g=1.2$ meV}
      \label{E_MES_05_10T}
\end{figure}

\begin{figure}[htbq]
      \includegraphics[width=0.40\textwidth,angle=0]{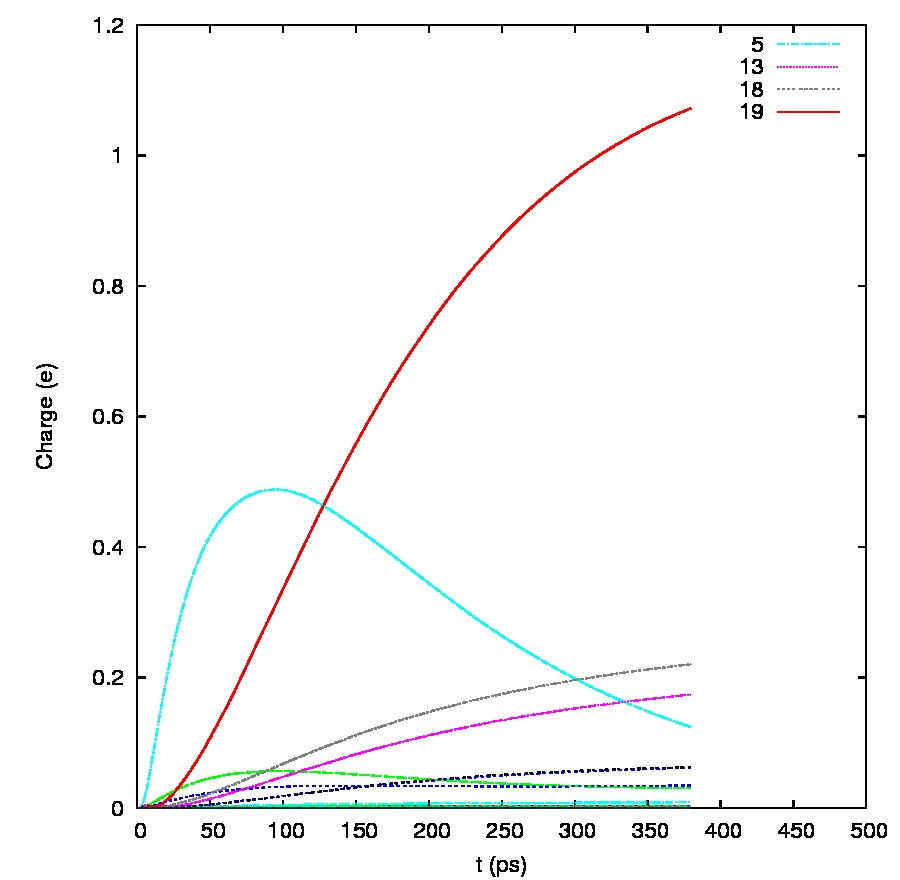}
      \includegraphics[width=0.40\textwidth,angle=0]{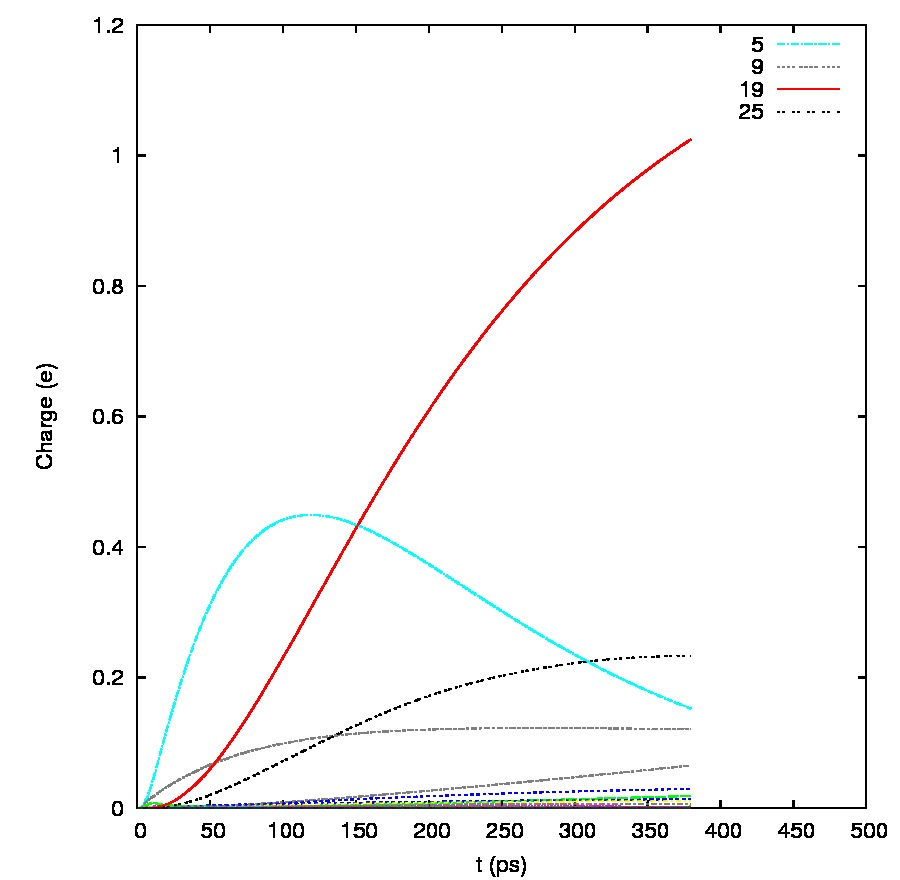}
      \caption{(Color online) For the interacting system the occupation of
              MESs as a function of time for $B=0.5$ T
              (upper panel), and $B=1.0$ T (lower panel).
              $L_x=300$ nm. At $B=1.0$ T $V_g=1.0$ meV, and at
              $B=0.5$ T $V_g=1.2$ meV. $\mu_R=1.6$ meV, $\Delta\mu =0.5$ meV, and $g_0^la_w^{3/2}=40$ meV.}
      \label{CorrC0510T}
\end{figure}

The lowest 32 levels of the many-electron energy spectra for the two different
values of the magnetic field are displayed in Fig.\ \ref{E_MES_05_10T},
where a care has been taken to identify the one- and two-electron MESs for both cases
with different symbols. Here, $\Delta\mu =0.5$ meV.
We note that for $B=1.0$ T the one- and two-electron MESs are
separated by a small energy gap, but not for $0.5$ T.
The time-dependent occupation probability of the MESs is
demonstrated in Fig.\ \ref{CorrC0510T}.
Further analysis of the state structure of the system is done in Table \ref{tab:B10} for
$B=1.0$ T, and in Table \ref{tab:B05} for $0.5$ T.
\begin{table}[htb]
\begin{center}
\begin{tabular}{lllr}
\hline
1e & $|5)$          & $E_5=1.28$ meV           & {}    \\
{} & $|9\rangle$    & $|000100.000000\rangle$   & 100.0\% \\
\hline
1e & $|9)$          & $E_9=1.88$ meV           & {}    \\
{} & $|129\rangle$  & $|000000.010000\rangle$   & 100.0\% \\
\hline
2e & $|19)$         & $E_{19}=3.36$ meV        & {}    \\
{} & $|6\rangle$    & $|101000.000000\rangle$   & 11.8\%\\
{} & $|11\rangle$   & $|010100.000000\rangle$   & 9.1\% \\
{} & $|13\rangle$   & $|001100.000000\rangle$   & 11.3\%\\
{} & $|18\rangle$   & $|100010.000000\rangle$   & 20.1\%\\
{} & $|25\rangle$   & $|000110.000000\rangle$   & 35.5\%\\
\hline
2e & $|25)$         & $E_{25}=3.68$ meV        & {}    \\
{} & $|34\rangle$   & $|100001.000000\rangle$   & 20.2\%\\
{} & $|41\rangle$   & $|000101.000000\rangle$   & 4.9\% \\
{} & $|66\rangle$   & $|100000.100000\rangle$   & 29.0\%  \\
{} & $|73\rangle$   & $|000100.100000\rangle$   & 12.3\%\\
{} & $|131\rangle$  & $|010000.010000\rangle$   & 17.8\%\\
\hline
\end{tabular}
\end{center}
\caption{The most probable interacting MES $|\mu )$ at $t=380$ ps
         and $B=1.0$ T together with their strongest components $|\nu\rangle$
         of non-interacting MESs. The Fock space representation of the
         states $|\nu\rangle$ is shown with a period indicating the location
         of the chemical potential $\mu_R$ in the right lead. $\Delta\mu =0.5$ meV, and $g_0^la_w^{3/2}=40$ meV.}
\label{tab:B10}
\end{table}
At $B=1.0$ T mainly the one-electron state $|5)$ is initially occupied with lesser
probability for $|9)$. Without the Coulomb interaction only these two states with the
same components will gain any significant probability of occupation. Similarly, as in
previous section where we analyzed the occupation of the system at a lower bias we
see here that the Coulomb interaction facilitates the coupling to the lower lying
states and thus increases the probability of the occupation of a two-electron state
with one of the electrons in a low energy SES. The next two-electron state to gain
significant occupation probability is $|25)$ with a bit higher energy and a
higher likelihood for one of the electrons to be just above $\mu_R$.

Very similar picture is seen in Table \ref{tab:B05} for the state-structure at $B=0.5$ T,
with the exception that only a single one-electron state gains significant occupation
probability and the two-electron states coming in after the most probable one, $|19)$,
are lower in energy than that one. Also, though not shown here the non-interacting
system at $B=0.5$ T will gain a slight probability for the occupation of a two-electron
states. This is caused by the missing gap between the one- and two-electron states that
we earlier pointed out in Fig.\ \ref{E_MES_05_10T}.
\begin{table}[htb]
\begin{center}
\begin{tabular}{lllr}
\hline
1e & $|5)$          & $E_5=1.31$ meV           & {}    \\
{} & $|9\rangle$    & $|00010.0000000\rangle$   & 100\% \\
\hline
2e & $|19)$         & $E_{19}=3.46$ meV        & {}    \\
{} & $|6\rangle$    & $|10100.0000000\rangle$   & 12.0\%\\
{} & $|11\rangle$   & $|01010.0000000\rangle$   & 22.4\% \\
{} & $|13\rangle$   & $|00110.0000000\rangle$   & 33.8\%\\
{} & $|19\rangle$   & $|01001.0000000\rangle$   & 4.6\% \\
{} & $|21\rangle$   & $|00101.0000000\rangle$   & 10.1\%\\
\hline
2e & $|13)$         & $E_{13}=3.02$ meV        & {}    \\
{} & $|4\rangle$    & $|11000.0000000\rangle$   & 88.2\%\\
{} & $|6\rangle$    & $|10100.0000000\rangle$   & 4.5\% \\
{} & $|13\rangle$   & $|00110.0000000\rangle$   & 6.4\%  \\
\hline
2e & $|18)$         & $E_{18}=3.33$ meV        & {}    \\
{} & $|7\rangle$    & $|01100.0000000\rangle$   & 10.4\%\\
{} & $|10\rangle$   & $|10010.0000000\rangle$   & 73.8\%\\
{} & $|37\rangle$   & $|00100.1000000\rangle$   & 11.3\%\\
\hline
\end{tabular}
\end{center}
\caption{The most probable interacting MES $|\mu )$ at $t=380$ ps
         and $B=0.5$ T together with their strongest components $|\nu\rangle$
         of non-interacting MESs. The Fock space representation of the
         states $|\nu\rangle$ is shown with a period indicating the location
         of the chemical potential $\mu_R$ in the right lead. $\Delta\mu =0.5$ meV, and $g_0^la_w^{3/2}=40$ meV.}
\label{tab:B05}
\end{table}

\begin{figure}[htbq]
      \includegraphics[width=0.42\textwidth,angle=0]{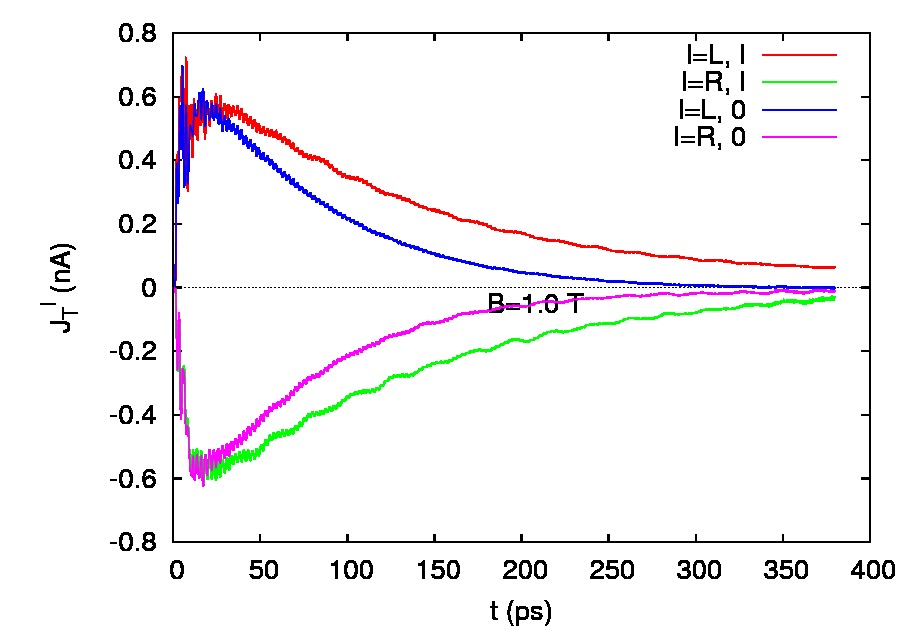}
      \includegraphics[width=0.42\textwidth,angle=0]{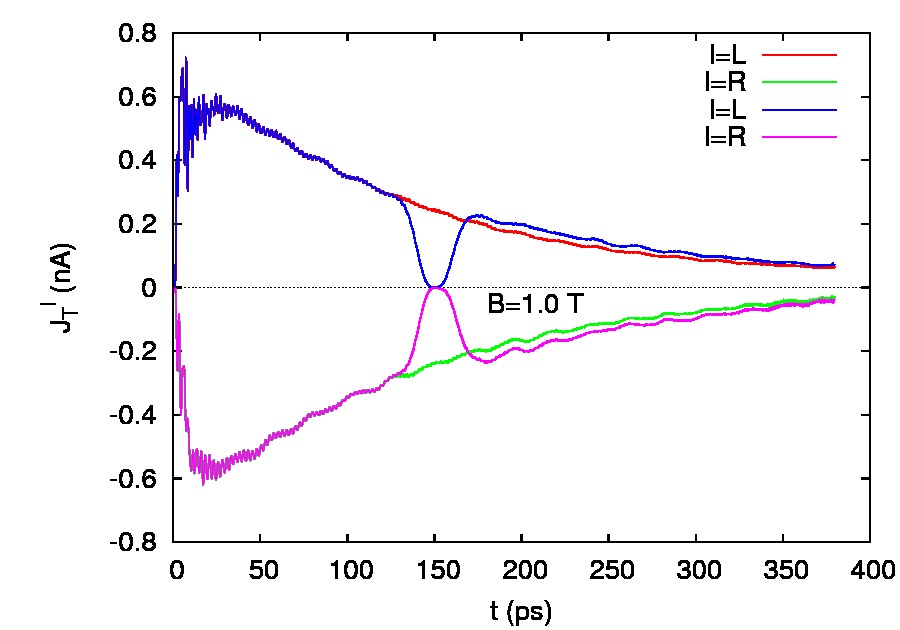}
      \includegraphics[width=0.42\textwidth,angle=0]{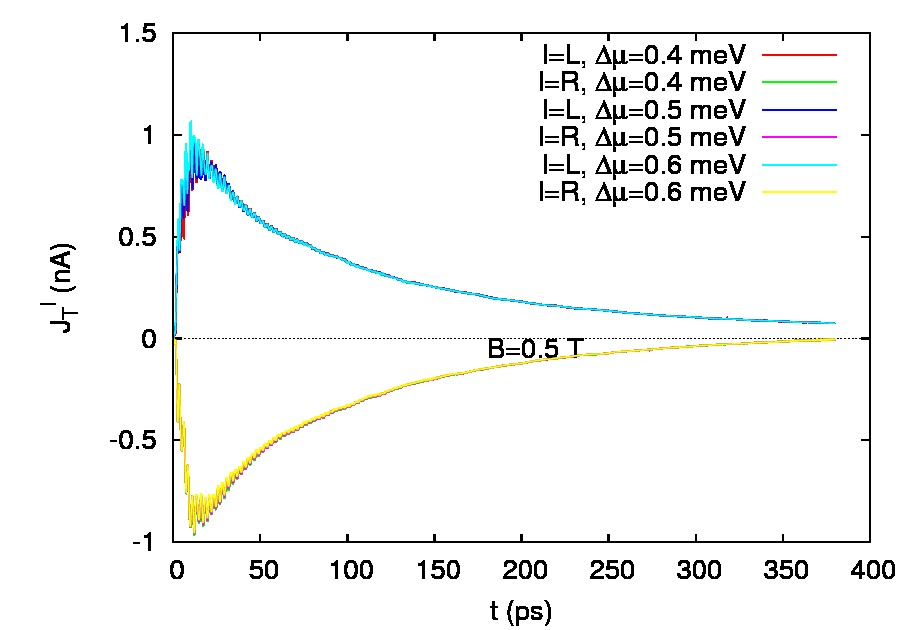}
      \caption{(Color online) The total current in the left and right leads as function of
               time for the non-interacting and the interacting system at
               $B=1.0$ T (top panel). The total current in the left and right leads
               for a system with steady coupling compared to a system where the coupling is
               momentarily switched-off smoothly around $t=150$ ps (center panel).
               The total currents for a system in $B=0.5$ T (bottom panel).
               $\Delta\mu =0.5$ meV, $L_x=300$ nm, and $g_0^la_w^{3/2}=40$ meV.}.
      \label{E_currCP0510T}
\end{figure}

Figure \ref{E_currCP0510T} compared the total current in the left and right leads for $B=1.0$ T
in the top panel for a non-interacting and an interacting system. Both show smooth oscillations
after the initial transient period, but they are clearer for the interacting system.
The center panel of Fig.\ \ref{E_currCP0510T} demonstrates that the oscillations in the
current are not changed to any extent by smoothly decoupling the system
momentarily from the leads around $t=150$ ps. The bottom panel of
Fig.\ \ref{E_currCP0510T} indicates that for the case of $B=0.5$ T the
oscillations are either absent or too weak to be discernible. Clearly, the system is
not in its ground state, the coupling to the leads moves it out of equilibrium,
and a glance at Tables \ref{tab:B10} and \ref{tab:B05} observing what kind
of states are available to the system awakens the question if the coupling has
generated collective oscillations as in the case of the closed ring subject to
strong external perturbation.\cite{Gudmundsson03:161301} In the case of the
system without an interaction the single electron states would have a restoring force
from the potential defining the wire and the ring which do not have a totally flat
bottom, see Fig.\ \ref{System}. But here the coupling to the leads is weak,
the momentary switch-off of it does not influence the oscillations, and an
inspection of the density confirms that the density only shows minute oscillations
that we will describe below.

\begin{figure}[htbq]
      \includegraphics[width=0.23\textwidth,angle=0,viewport=12 12 249 250,clip]{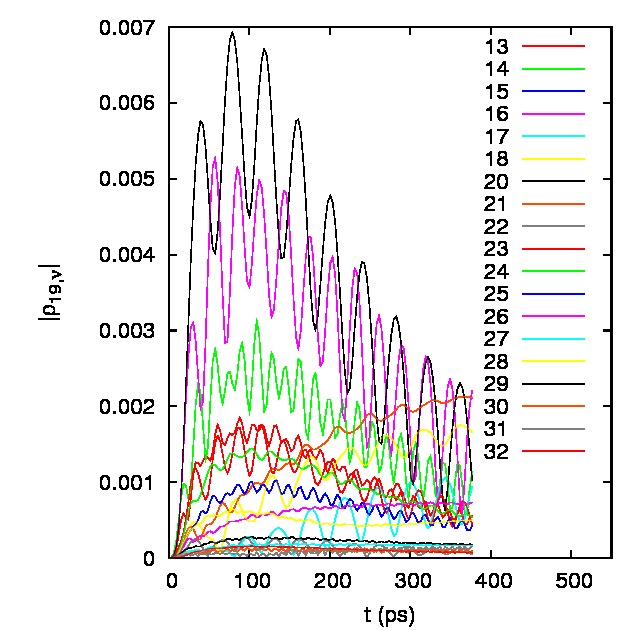}
      \includegraphics[width=0.23\textwidth,angle=0,viewport=12 12 249 250,clip]{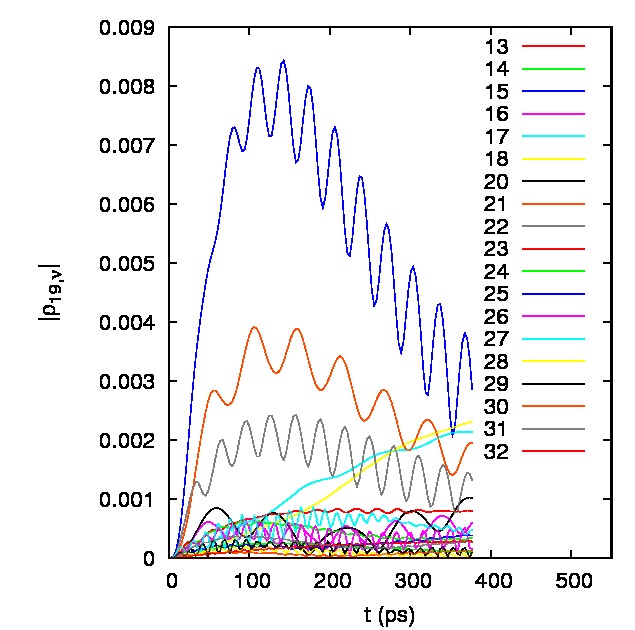}
      \caption{(Color online) The strength of the off-diagonal element $|\rho_{19,\nu}|$
               as a function of time for $B=0.5$ T
               (left panel), and $B=1.0$ T (right panel).
               $L_x=300$ nm. At $B=1.0$ T $V_g=1.0$ meV, and at
               $B=0.5$ T $V_g=1.2$ meV. $g_0^la_w^{3/2}=30$ meV.}
      \label{RDO_19n}
\end{figure}

The dynamic evolution is governed by the GME (\ref{GME}) and in Fig.\ \ref{RDO_19n}
we display the correlation of the two-electron state $|19\rangle$ that gains the
highest occupation probability with time in our system, i.e.\ the off-diagonal
elements of the RDO, $\rho_{19,\nu}$.
For both values of the magnetic field we see indeed oscillations with comparable
period as the smooth oscillations in the current, but in the case of the lower
magnetic field many elements show strong oscillations but not in phase.
For the higher magnetic field only one or two elements oscillate and one of them
is clearly stronger.

We see thus that oscillations of the electron correlations are inherent in the
GME formalism irrespective of the presence of the Coulomb interaction or not.
It is a part of the correlations forced on various states of the system by the coupling to the
leads. Here, we observe that the Coulomb interaction further couples different
types of states in the system. States with weak coupling to the leads residing
in regions of the system away from the contact area with states with higher
presence in the contact area. Furthermore, the magnetic field simplifies the
energy spectrum of the system such that oscillations in the correlation
of a single pair of MESs will be dominant and thus visible in the total
current.

\begin{figure}[htbq]
      \includegraphics[width=0.23\textwidth,angle=0,viewport=12 12 214 214,clip]{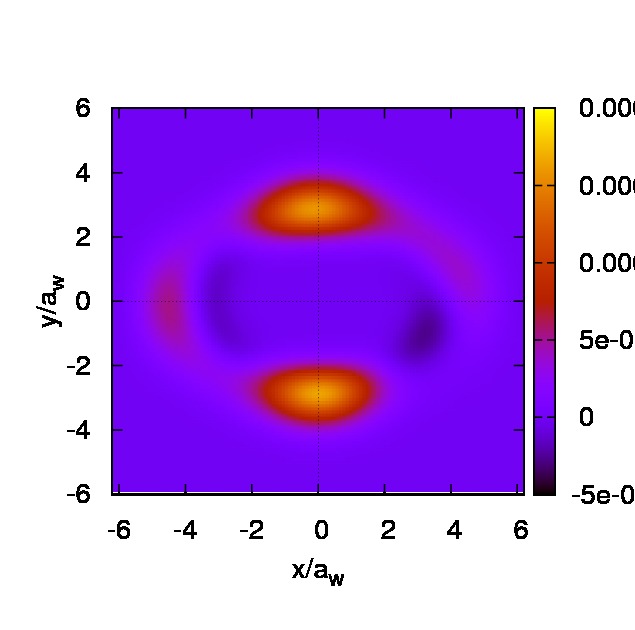}
      \includegraphics[width=0.23\textwidth,angle=0,viewport=12 12 214 214,clip]{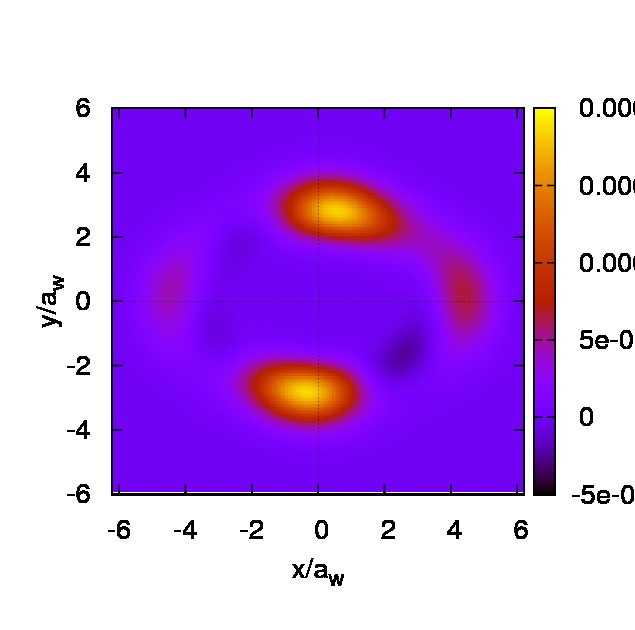}
      \includegraphics[width=0.23\textwidth,angle=0,viewport=12 12 214 214,clip]{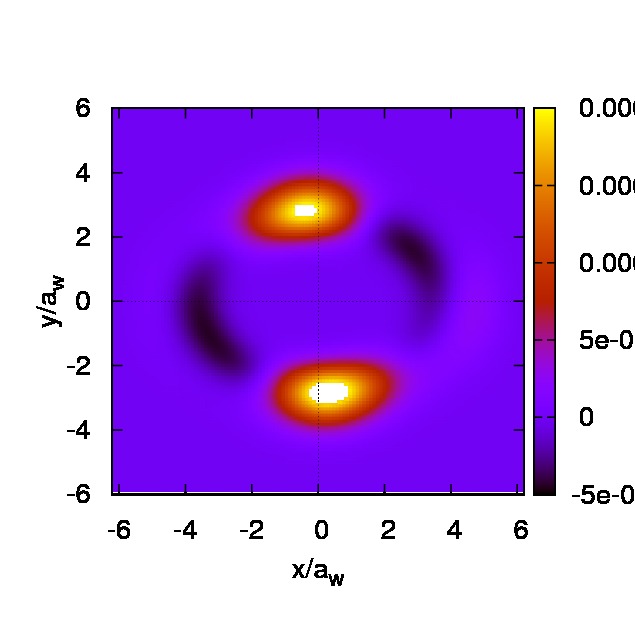}
      \includegraphics[width=0.23\textwidth,angle=0,viewport=12 12 214 214,clip]{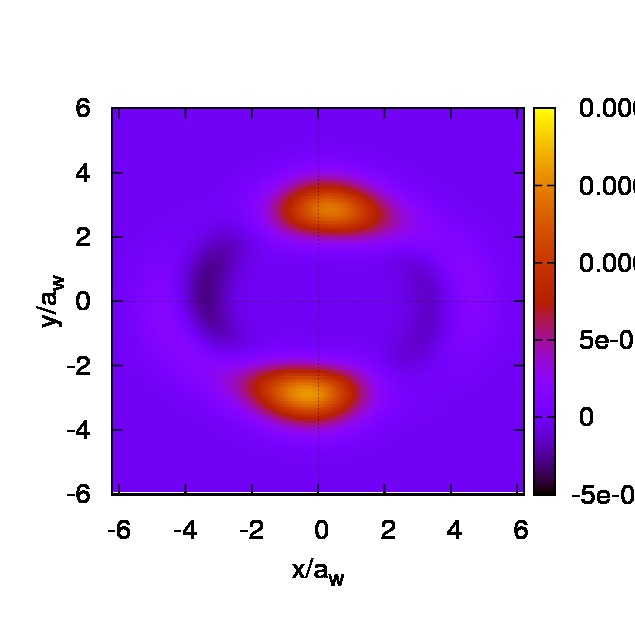}
      \caption{(Color online) The induced density $n(\mathbf{r},t)-n(\mathbf{r},t-\delta t)$
              at time points, $252.2$ ps (upper left panel), $264.3$ ps
              (upper right panel), $273,4$ ps (lower left panel), and $282.5$ ps
              (lower right panel). $B=1.0$ T, and $g_0^la_w^{3/2}=40$ meV.}
      \label{Qnn_C510T}
\end{figure}

As was stated before the oscillations in the density caused by the oscillation
in the electron correlation are minute. We thus display in Fig.\ \ref{Qnn_C510T}
the ``derivative'' density or induced density defined by
$n(\mathbf{r},t)-n(\mathbf{r},t-\delta t)$, where we have taken $\delta t=1.5$ ps
instead of comparing always to the density at a certain fixed point in time, the
reason being that the electron charge is still increasing in the time
interval used and we see the density peaks in the ring away from the
contact area always growing.

In addition, we see that electron density or charge is shifted between the peaks and the
contact area thus influencing the effective coupling between the system and the leads.

In order to check the stability of the results with respect to the exact location of the
window of chemical potential we have repeated the calculations for $\mu_R=2.0$ meV.
The results are very similar with respect to the oscillations observed in the current
and the occupation of the MESs. The states gaining highest occupation are still the same
but at $\mu_R=2.0$ meV additional MESs with a bit higher energy show up with low occupation
that were almost empty for $\mu_R=1.6$ meV. At present we do not feel confident to increase
the height of the bias window further due to the fact that we have only included 12 SESs in
the calculation.

%
%
\section{Summary}
We have used a time-dependent transport formalism built on the generalized master equation
(GME) where the mutual Coulomb interaction between electrons is treated within
the ``exact numerical diagonalization'' or ``configuration interaction'' to analyze
the transport properties of a quantum two-dimensional ring or a short 2D quantum wire
with an embedded ring. The quantum wire is defined by a parabolic confinement potential
in the wire plane perpendicular to the transport direction. The ends of the quantum
wire are hard walls that are made transparent to tunneling by a weak coupling to leads.
The shape of the finite wire and the definition of the ring potential (\ref{V_QR})
leads to a ring system with a potential that does not have a totally flat bottom,
see Fig.\ \ref{System}. For this reason we have a system that has SESs that are
ring states for higher energy, but the lowest states can be slightly localized in different
part of the system. We show that indeed, in this geometry the Coulomb interaction between
the electrons increases the occupation of the system by coupling states with different
localization properties. The correlation of the states in the system caused by the
coupling to the leads is enhanced by the Coulomb interaction leading to a behavior that
runs counter to the usual Coulomb blockade in a simpler geometry. Of course the
Coulomb blocking mechanism is inherent in the interacting system, but by comparing the
interacting system with the noninteracting one we discover finer details in the
action of the interaction, details that are usually collected under the title:
correlation effects.

In addition we find current oscillations that are caused by oscillating correlation
properties of the electrons in the system. These oscillations become visible in
higher magnetic field due to the reduction of MESs active in the transport by the
magnetic field.
As the correlations are caused by the coupling to the leads the oscillations are
visible in systems without or with Coulomb interaction between the electrons, but
the Coulomb interaction influences them through its enhancing of correlations by
coupling of electronic states.

\begin{acknowledgments}
      The authors acknowledge financial support from the Research
      and Instruments Funds of the Icelandic State,
      the Research Fund of the University of Iceland, the
      Icelandic Science and Technology Research Programme for
      Postgenomic Biomedicine, Nanoscience and Nanotechnology, the
      National Science Council of Taiwan under contract
      No.\ NSC97-2112-M-239-003-MY3, and the Reykjavik University Development
      Fund T09001. V. M. acknowledges the hospitality of the Reykjavik University, Science Institute
      and the partial support from PNCDI2 program (grant No. 515/2009 and Grant No. 45N/2009).

\end{acknowledgments}
%
%
%
\bibliographystyle{apsrev}

%
%
%
\end{document}